# BOLD-fMRI in the mouse auditory pathway

Guilherme Blazquez Freches†, Cristina Chavarrias†, Noam Shemesh*

*Champalimaud Neuroscience Programme, Champalimaud Centre for the Unknown, Lisbon, Portugal*

Abbreviated title: Mouse auditory pathway revealed by BOLD fMRI

**Conflict of Interest**: None

**Acknowledgements:** This study was funded in part by the European Research Council (ERC) under the European Union's Horizon 2020 research and innovation programme (grant agreement No. 679058 - DIRECT-fMRI).

*Corresponding author
Dr. Noam Shemesh, Champalimaud Neuroscience Programme, Champalimaud Centre for the Unknown
Av. Brasilia 1400-038, Lisbon, Portugal
E-mail: noam.shemesh@neuro.fchampalimaud.org; Phone number: +351 210 480 000 ext. #4467

†These authors contributed equally to this paper

# Abstract

The auditory pathway is widely distributed throughout the brain, and is perhaps one of the most interesting networks in the context of neuroplasticity. Accurate mapping of neural activity in the entire pathway, preferably noninvasively, and with high resolution, could be instrumental for understanding such longitudinal processes. Functional magnetic resonance imaging (fMRI) has clear advantages for such characterizations, as it is noninvasive, provides relatively high spatial resolution and lends itself for repetitive studies, albeit relying on an indirect neurovascular coupling to deliver its information. Indeed, fMRI has been previously used to characterize the auditory pathway in humans and in rats. In the mouse, however, the auditory pathway has insofar only been mapped using manganese-enhanced MRI. Here, we describe a novel setup specifically designed for high-resolution mapping of the mouse auditory pathway using high-field fMRI. Robust and consistent Blood-Oxygenation-Level-Dependent (BOLD) responses were documented along nearly the entire auditory pathway, from the cochlear nucleus (CN), through the superior olivary complex (SOC), nuclei of the lateral lemniscus (LL), inferior colliculus (IC) and the medial geniculate body (MGB). By contrast, clear BOLD responses were not observed in auditory cortex (AC) in this study. Diverse BOLD latencies were mapped ROI- and pixel-wise using coherence analysis, evidencing different averaged BOLD time courses in different auditory centers. Some degree of tonotopy was identified in the IC, SOC, and MGB in the pooled dataset though it could not be assessed per subject due to a lack of statistical power. Given the importance of the mouse model in plasticity studies, animal models, and optogenetics, and fMRI's potential to map dynamic responses to specific cues, this first fMRI study of the mouse auditory pathway paves the way for future longitudinal studies studying brain-wide auditory-related activity *in vivo*.

**Graphical Abstract**

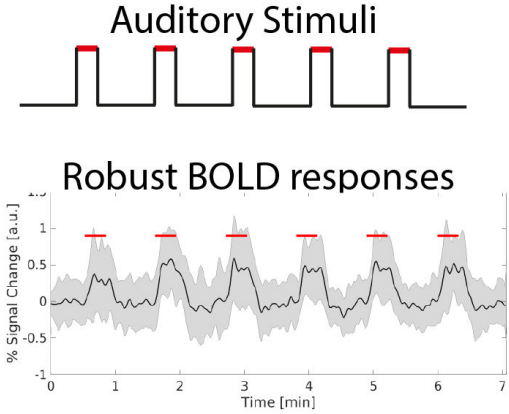
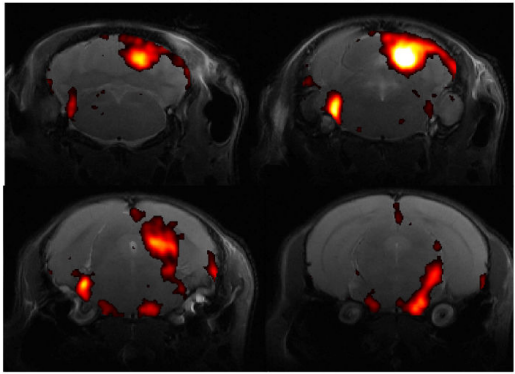

# Introduction

The auditory system evolved as an essential tool for survival across a wide range of species. It enables a finely-tuned deconstruction of a specific sound into its basic characteristics, e.g., frequency and amplitude, and it can adapt to extreme situations through neuroplasticity in auditory receptors as well as throughout the auditory pathway. Such flexibility persists even in adulthood: in rodent models, for example, neuroplasticity is evident from the complete recovery occurring after deep injury (Chambers et al., 2016) or from adaptation to adverse conditions (Lau et al., 2015). However, the mechanisms underlying both activity and plasticity in the auditory pathway requires the design of methods capable of characterizing the entire pathway *in vivo* and longitudinally.

Magnetic Resonance Imaging (MRI) is one of the most powerful noninvasive techniques for probing global brain function *in vivo*. Functional MRI (fMRI), in particular, offers the possibility of investigating global brain function in a longitudinal fashion, albeit through a somewhat indirect mechanism – the Blood-Oxygenation-Level-Dependent (BOLD) contrast (Ogawa et al. 1990) – which can be considered a surrogate reporter of underlying neural activity (Logothetis, 2008). The hemodynamic response associated with BOLD is known to vary across brain regions with complex dynamics arising either from physiological factors such as the geometry of the capillary bed, or from different patterns of neural activation for regions involved in executing a task (John T. Cacioppo, Louis G. Tassinary, 2000). Thus, despite its indirect nature, BOLD can still be very useful for mapping spatial locations of neural activity, and in some cases, also reflect some of its dynamics. In rodents, BOLD is typically easier to detect in rats than mice using fMRI (Jonckers et al., 2015), and fMRI in the rat was indeed widely used for studying disorders such as stroke (Canazza et al., 2014), epilepsy (Blumenfeld, 2007), ischemia (Chan et al., 2010), or to study plasticity (Lau et al., 2015), e.g., after nerve deafferentation. Global brain function arising from tactile stimulation (fore/hind paws, whiskers) was extensively studied in normal animals (Shih et al., 2013); in addition, activation patterns arising from visual (Bailey et al., 2013; Lau et al., 2011a, 2011c; Matthew C. Murphy et al., 2016; Niranjan et al., 2016; Pawela et al., 2008; Tambalo et al., 2015; Van Camp et al., 2006), olfactory (Matthew C Murphy et al., 2016) and, recently, auditory stimuli (Cheung et al., 2012a) were reported. The latter, in particular, was only recently introduced, presumably due to the difficulty in delivering pure tones in a very loud environment of an MRI scanner. Nevertheless, introducing auditory stimuli to rats was very worthwhile, as these stimuli have a very rich dimension space (e.g. frequency, amplitude, timing, timbre) and indeed, the above-mentioned studies have already led to much insight into the auditory pathway. For example, using specialized MRI-compatible hardware, the auditory circuit was mapped in its entirety, and very precise tonotopy in the Inferior Colliculus (IC) was demonstrated (Cheung et al., 2012b). Sound pressure level encoding in the IC (Zhang et al., 2013), the interaural difference encoding in the auditory system (Lau et al., 2013), the influence of the IC in deviant sound detection (Gao et al., 2014), ultrahigh frequency encoding in the IC (Gao et al., 2015a), modulations of AC and visual cortex on IC processing (Gao et al., 2015b), and the effects of long term exposure to certain types of sounds (Lau et al., 2015), were also very recently reported by the same group in the rat, showing the utility of the fMRI approach of investigating the system in its entirety. Despite these impressive rat studies, fMRI in the mouse has not been heretofore performed, likely due to its small brain, that requires high resolution which in turn leads to sensitivity issues and perhaps since mice can be relatively unstable under sedation, and their BOLD responses can become uncoupled from the neural activity under anesthesia (Masamoto and Kanno, 2012; Schlegel et al., 2015a). To avoid these onuses, plasticity in the mouse auditory pathway has been studied by Turnbull's group through manganese enhanced MRI (MEMRI), which enabled the mapping of sound-evoked activity following several hours of exposure (Yu et al., 2008, 2007, 2005). However, it is clear that functional MRI-based mapping could potentially provide different information than MEMRI, namely, by mapping specific BOLD responses following direct auditory stimulation, their onset and dynamics, and, for example, how they adapt to different stimuli. Furthermore, the mouse's prominence in transgenesis as well as in multiple models for hearing deficits and plasticity (Polley et al., 2006), make fMRI in the mouse auditory pathway an attractive goal.

On the other hand, since its inception, ROI coherence analysis has been a tool devoted to ascertain functional connectivity for in a vast number of brain networks including the auditory system (Schlee et al., 2008; Sun et al., 2004). The goal of coherence analysis is to compute correlations of time series with respect to a reference seed time-series. This allows for the identification of task-specific patterns across brain regions without making any assumptions about the brain response shape in those centers (Lindquist, 2008), assuming a linear system.

Here, we establish a robust system for auditory pathway mapping in the sedated mouse. Using high-field fMRI, we show, for the first time, the salient characteristics of most of the mouse auditory pathway and the ensuing BOLD profiles *in vivo*. We find robust BOLD responses along the pathway, from which active regions can be mapped, providing clues to their connectivity; evidence for distinct delays in the hemodynamic responses across auditory centers is revealed from coherence analyses of the BOLD signals. These experiments thus pave the way for longitudinal global brain characterization of auditory pathway activity, plasticity, and aberration in vivo.

# 1 Materials and Methods

All experiments were preapproved by the Champalimaud Centre for the Unknown's Ethics Committee. Adult C57bl mice (N=5 males and N=7 females) weighing ~25 g and aged between 7 to 8 weeks old housed under ad libitum conditions were used for the auditory fMRI experiments.

## 1.1 Setup for delivery of auditory stimulus in the scanner

The setup shown in Figure 1A was designed to deliver precise auditory stimuli within the magnet. A 2-channel soundboard (YAMAHA AG-03, 10-1, Nakazawacho, Naka-ku Hamamatsu-shi, Shizuoka, 430-8650, Japan) with a dynamic range of 24 bits and 2.451 $V_{RMS}$ output before clipping was used to sample the wide range of frequencies relevant for mouse audition (up to 96 kHz with a sampling rate of 192 kHz). An in-house designed voltage amplifier capable of a 3x amplification up to 24 Vpp was used for amplifying the signal to the levels required by ultrasonic speakers. To facilitate robust performance, a piezoelectric speaker (L010 KEMO, Leher Landstr. 20 D-27607 Geestland, Germany) previously shown to produce ultrasonic sounds at high output levels (+100 dB) with a relatively flat frequency response up to 75 kHz (Cheung et al., 2012a) was used to produce the sound waves. Finally, guiding these sound waves into the mouse ear in an efficient manner (i.e., allowing for optimal placement of the mouse in the cryocoil (Figure 1A) required specialized tubing of different diameter. The interface between the speaker and the main tube (made with nylon and with an inner diameter of 14 mm) was accomplished by a pair of custom 3D-printed connectors made of polyethylene terephthalate glycol. At the end of the main tube, the circuit had to be tapered down to the width of the final semi-rigid tube that inserts into the mouse ear; a screw lock mechanism secured the coupling of the main tube and the tube that reached the ears (PETG, 3.2 mm diameter). The total length of the tubing was 88 cm (including 6 cm of Tygon tubing). Interfacing with the scanner was accomplished using an Arduino microcontroller (ARDUINO, Officine Arduino/Fablab Torino Via Egeo 16 10134 Turin, Italy) as a trigger detector; the triggers were sent by programming trigger lines into pulse sequences, which then produced triggers when a sound cue was due.

Sound spectra were recorded 30 cm from the magnet bore with a Brüel & Kjær 4939-A-011 - ¼-inch free-field microphone with Type 2670 preamplifier, 4 Hz to 100 kHz, 200 V polarization (Skodsborgvej 307 DK-2850 Nærum Denmark).

## 1.2 MRI experiments

### 1.2.1 General.

All MRI experiments were performed on a 9.4 T Bruker BioSpec scanner (Karlsruhe, Germany), equipped with a gradient system capable of producing up to 660 mT/m in all directions. An 86 mm quadrature resonator was used for transmission, while a 4-channel array cryogenic coil (Bruker BioSpin, Karlsruhe, Germany) was used for reception. In all experiments, temperature and respiration sensors were placed for monitoring mouse physiology (Model 1030 Monitoring Gating System, SA Instruments, United States of America), and heating was achieved by warm water circulation. Offline experiments utilized transcutaneous $pCO_2$ measurements to further ensure physiological stability during the experiments.

*1.2.2 Anesthesia and preparation for MRI experiments*

Twelve mice were anesthetized briefly with 4% isofluorane (VIRBAC, Carros cedex, France) mixed with ambient air maintained by a vaporizer (VETEQUIP, Livermore, CA United States) in a custom-built box. The isofluorane concentration was reduced to ~3% after ~3 minutes, and the animals were quickly moved to the animal bed and stabilized with a nose cone and a bite bar. A subdermal bolus of medetomidine at 0.4 mg/kg was injected to the mice, followed by a constant infusion of 0.8 mg/kg/h (Adamczak et al., 2010) delivered by a syringe pump (GenieTouch, KentScientific, Connecticut, USA) through the same route. Isoflurane was reduced at a rate of 1% every 6 minutes until it reached 0%, and oxygen levels were carefully maintained at 27% with the aid of an oxygen sensor (Viamed, Cross Hills, United Kingdom).

Once sedated, the flexible Tygon tube (Figure 1A) was carefully inserted into the mouse ear. The tube was held in place by occluding both mice ears with liquid (50 °C) Paraffin (Sasolwax, Hamburg, Germany). This also served to isolate the mouse from external sounds. The entire assembly was then inserted to the scanner.

*1.2.3 MRI scans*

Once the animal was properly positioned in the scanner, scout images were acquired to assess the quality of position and to avoid strong tilts of the head due to the sound delivery system. In case the head was tilted (judged visually), animals were removed from the scanner and repositioned such that they were as aligned in the scanner as possible.

A $B_0$ field map was obtained from phase images acquired using a 3D dual-gradient-echo pulse sequence. The correction of $B_0$ field inhomogeneities was automatically performed using the MAPSHIM routine in a cuboid volume comprising ~525 mm$^3$ located in the brain and centered on the middle slice of the fMRI acquisitions.

An anatomical reference scan was acquired using a Turbo Rapid Acquisition with Relaxation Enhancement (RARE) sequence, with the following parameters: RARE factor = 8, Number of averages: 6, TR/TE = 2000/10 ms, FOV = 15 x 15 mm$^2$, data matrix = 200 x 200, with six parallel 0.650 mm thick slices, separated by gaps of 0.150 mm (approximately Bregma -5.2 mm to -2 mm). In all experiments, the breathing rate and rectal temperature were monitored throughout the experiment.

### 1.2.4  fMRI paradigm and sequence parameters

The fMRI experiments were performed using the fast-imaging with steady-state-precession (FISP) pulse sequence refocusing all coherence pathways (trueFISP), which was chosen due to its excellent SNR, marked $T_2/T_1$ contrast, high spatiotemporal resolution, and robustness against image distortions (Park et al., 2011; Zhou et al., 2012). With the gain of a factor of 2-3 in signal from the cryoprobe, highly resolved images could be acquired using the following parameters: TR/TE = 2.8/1.4 ms, flip angle 45°, six slices of 650 µm thickness, FOV = 15 x 15 mm$^2$, matrix = 100 x 100 leading to an in-plane resolution of 150 x 150 µm$^2$. The effective repetition time for the entire volume was 1.307 s, and the slice acquisition order was interleaved. The slices were acquired in the same position as in the anatomical scans.

### 1.2.5  Experiments description

fMRI experiments were performed using the block paradigm shown in Figure 2A (upper panel), each block encompassing 34 images at rest (~44 s) and 16 images (~21 s) at auditory stimulation. This basic block was repeated 6 times to complete a fMRI run, (total number of volumes acquired of 334, lasting 7m16s per run). Therefore, a single pure tone was delivered six times per run, of either 5, 12 or 20 kHz. The mice were then allowed to rest for at least 5 minutes before another run (with a different pure tone from the former run) was performed. Between 3-6 such runs were completed for each mouse in every session, depending on the animal's physiological stability, giving a total number of n=18 runs for the combined analysis.

Given the high SNR provided by the cryogenic coil, we also sought to explore whether closely-spaced tones generated discernible patterns in fMRI signals. We performed an additional experiment on n = 7 animals with 1 kHz separation between tones, as shown in Table 1. To avoid adaptation, the respective experiments harnessed the block paradigm shown in Figure 2A (bottom panel). Each block again encompassed 34 images (~44 s) at rest and 16 images (~21 s) with auditory stimulus, but now in each block, a different pure tone randomly drawn from 35,36,37,38 or 39 kHz was delivered. These randomized stimuli are depicted in different colors in Figure 2A. Six of these stimulation blocks were presented per fMRI run, again consisting of 334 volumes acquired during 7m16s. At least 5 minutes were again allowed to elapse between runs, and between 3-6 runs were completed for each mouse scanned, giving a total number of n = 26 runs.

In total, each experimental session lasted about 1.5 h - approximately 30 min for animal preparation, 30 min more for acquisition of preparatory MRI scans, and the remaining 30 minutes were spent acquiring the fMRI data. The animals typically started moving and waking up after around 1.5 h. The total numbers of animals, runs and stimulation description are shown in Table 1. Following the experiments, atipamezole hydrochloride (SEDASTOP, AnimalCare, York, United

Kingdom) at 0.1 mg/kg administration was injected to allow full recovery from medetomidine sedation.

## 1.3 Data Analysis

### 1.3.1 Preprocessing

Images were converted to Nifti, corrected for slice timing, realigned to mean, normalized (rigid plus affine transformations) and smoothed (3D kernel with FWHM = 0.3 mm) before GLM fitting (Friston et al., 2007). All these preprocessing steps used functions from SPM12 (Wellcome Trust Centre for Neuroimaging, London, UK) and fMRat (Chavarrias et al., 2016). These preprocessed images were the input for both GLM statistical mapping and coherence analyses.

### 1.3.2 ROI analysis

Regions along the auditory pathway were delineated in ImageJ (Schneider et al., 2012) according to the Paxinos and Franklin atlas (Paxinos and Franklin, 2004), converted to binary, and applied to the preprocessed datasets as well as the averaged datasets. Their delineation is shown in supplementary Figure S1, where all regions are tagged with their atlas abbreviation followed by their slice number: CN=cochlear nucleus, IC=inferior colliculus, LL=lateral lemniscus, SOC_IL=ipsilateral superior olivary complex, SOC_CL=contralateral superior olivary complex, MGB=medial geniculate body, AC=auditory cortex. The time-courses were detrended and temporally filtered at 0.95 of the Nyquist frequency (363 mHz). These corrected time-courses, as well as their averaged cycle, were plotted, and the former were used for ROI coherence analysis with the fastest region exhibiting sufficient BOLD signal, CN1 (cochlear nucleus in slice 1), chosen as the seed (see more details below in Section 2.3.4).

*1.3.3   GLM analyses*

For the statistical t-maps, different GLMs were fitted, targeting (1) a per-subject analysis and (2) a group-level analysis. In all cases each session from every animal was regressed with their respective realignment parameters and a single gamma function peaking at 2.5 s (SPM function spm_hrf() with input vector p = [4.5 8 2 1 Inf 0 32]) was used as the haemodynamic response. The first 10 volumes were skipped because the steady-state in the FISP was only reached approximately at volume 5. For both global and per-subject analyses the appropriate SPM contrasts were built in order to obtain maps for the global auditory response as well as for each individual frequency.

*1.3.4   Coherence analysis*

In this study, two types of coherence analysis were performed: ROI-based (Ashby, 2011), and voxel-by-voxel (all coherence analyses were generated from the preprocessed images using customized Matlab® (The Mathworks, Nattick, MA, USA) scripts). For the ROI-based analysis, each ROI (defined in Figure S1) was used as the seed voxel against each other ROI, and, Matlab's mscohere and cpsd functions were used to generate coherence maps and coherence phase, respectively. The delay times were then computed from the slope of the unwrapped coherence phase associated with the first (lowest) 5 frequencies, where coherence was highest. For the voxel-by-voxel coherence mapping, the exact same procedure was repeated but now in the aim of generating detailed delay maps. The most caudal CN ROI (region CN1, Figure S1, first slice) was used as the seed, and its time-course extracted from the global averaged dataset. Its coherence with each voxel was then computed voxelwise. A mask was then created by thresholding the coherence amplitude maps above 0.4, and the mask was applied on the coherence phase maps. Finally, the delays were calculated in the masked regions from the coherence phase (Ashby, 2011), using the exact same procedure described above for the ROIs. Coherence (to CN1) amplitude and delay estimation were computed both for the averaged dataset and for each individual subject.

### 1.3.5 Tonotopic maps

Tonotopic maps were generated by thresholding each frequency's respective GLM positive t-map to voxels with the 3 % highest t- values in the particular slice analyzed, and then the different t-maps corresponding to different frequencies were overlaid on the anatomical reference scan using MRIcron (Rorden and Brett, 2000). Each frequency was color-coded, thus delineating the regions associated with the highest activation for that particular frequency. The different colors are represented with a 60% transparency to allow the visualization of the overlapping regions. This representation was performed for both the group map (all subjects pooled together) as well as for each independent subject in Experiment 1 and Experiment 2.

# 2 Results

A preceding step to auditory pathway mapping is the assessment of the sound quality entering the mouse's ear and potential interferences with scanner noises. Figure 1B shows a spectral analysis of all auditory stimuli used in this study, as well as the spectrum of scanner noise/sounds during the fMRI experiment (measured in separate occasions). All pure tones exhibited sharp and distinct maxima, always occurring at sound pressure levels > 6 dB higher than the background noise or the respective harmonics. The trueFISP sequence used for fMRI acquisition shows a broad spectral energy dispersion with prominent peaks, especially at lower frequencies, but even these are again always at least 6 dB lower than the delivered auditory stimuli. Importantly, the envelope of trueFISP pulse sequence is constant, leading to even less interference with the sounds delivered.

The robustness and reliability of the raw data is shown in Figure 2, which presents raw data (not averaged, not smoothed) from a single run in a single animal. The raw FISP images corresponding to a single run of the paradigm, in a single animal, are shown in Figure 2B. Signal to noise in the individual images was excellent, approaching 50 in the worst-case-scenario. ROIs depicting the raw signal time course for three prominent regions in the auditory pathway are also shown for a single run in a single representative mouse (Figure 2C). Interestingly, BOLD responses could be observed in the relevant ROIs even with the naked eye in single runs, even before averaging on the multiple experiments or on different animals.

## 2.1 ROI analysis

To gain insight into specific BOLD patterns and characteristics in the auditory pathway, the data from all animals, as well as the averaged stimulation cycle BOLD signals, were plotted for each of the ROIs along the pathway (six ROIs are shown in Figure 3, and all other ROIs are plotted in supplementary Figure S2; the ROI delineation is shown in supplementary Figure S1).

As observed in Figure 3, the time-courses show highly consistent activation in most ROIs, with approximately 0.5 % signal change for all cochlear nucleus (CN1, CN2, CN3), inferior colliculus (IC1, IC2, IC3), the superior olivary complex in slice 4, both ipsilateral (SOC_IL4) and contralateral (SOC_CL4), and in slice 3 only contralateral (SOC_CL3). Lateral lemniscus (LL3 and LL4), medial geniculate body (MGB4 and MGB5) and the ipsilateral superior olivary complex in slice 3 (SOC_IL3) exhibit approximately 0.2-0.4% signal changes. In these experiments, the auditory cortex (AC4 and AC5) evidences no signal contrast, and the full time-course exhibits only a very low-frequency oscillation.

Averaged BOLD characteristics can also be quantified in terms of, for example, time-to-peak. Most of the regions reach their maximum contrast about 8 s from the stimulus onset (CN1, IC1, IC2, IC3, SOC_IL4, LL3, LL4, MGB4 and MGB5), though there are some interesting exceptions. To guide the eye, a vertical line at t = 10 s from the stimulation onset has been overlaid on all averaged cycle plots. In slice 4, the contralateral SOC shows a clearly slower increase in BOLD as compared to the ipsilateral SOC in the same slice (Figure 3, SOC_IL4 and SOC_CL4), and the same applies to both ipsi- and contralateral-SOC in slice 3 (Figure S2). Also, the cochlear nucleus regions on the more rostral slices show a slower increase in BOLD signal as compared to the rest of the nuclei (Figure S2, CN2 and CN3).

## 2.2  Global auditory response

We sought to examine whether such BOLD responses could form the basis for auditory pathway mapping *in vivo*. A GLM analysis of the global auditory response including all 44 runs across 12 animals was performed and compared with its homologous coherence amplitude map. Highly significant BOLD responses (uncorrected p<0.001, t>4.40) were observed in the vast majority of the auditory pathway's crucial junctions (Figure 3) and approximately the same regions exhibit a strong coherence to CN1. Highly significant BOLD responses were documented in CN, SOC (bilaterally), ML, LL, IC, and MGB, with the most significant activation detected in CN and in IC. Interestingly, BOLD responses were not detected in the AC, neither by the conventional GLM analysis nor the thresholded coherence amplitude. While the highest BOLD signal is located in the inferior colliculus, the highest coherence to CN1 is located in the cochlear nucleus, as expected.

The group-level maps for each frequency (5 kHz, 12 kHz, 20 kHz, 35-39 kHz) are shown in Figure 5. These global, multisubject auditory maps for each pure tone show significant activation across all the auditory pathway, though with different significance levels. Only the map corresponding to tones of 20 kHz fails to surpass the statistical threshold in some regions like MGB and SOC. The highest t-values in the IC are located more medial for the highest frequency pure tones and more dorsal for the lowest tones. The coherence to CN1 is shown in panel B, and it highlights nearly the same regions as the GLM analysis, perhaps adding a few smaller and scattered clusters. For the sake of a fair comparison these maps were not thresholded by cluster size.

## 2.3 ROI-based Coherence analysis

Since the intrinsic BOLD characteristics, and in particular, the exact nature of neurovascular couplings in different regions along the pathway are unknown, it is instructive to harness a different, data-driven approach that is not as dependent on choice of specific model for intrinsic hemodynamics response profiles. Figure 6 shows coherence phase - a metric considered uncoupled to the intrinsic BOLD dynamics (Sun et al., 2004) between the CN1 ROI (chosen as the seed since it is first synaptic relay for neural activity in the circuit) and each other ROI (for ROI definitions see Figure S1). Importantly, assuming linearity in the neural responses, the coherence phase allows to extract time delays corresponding to temporal lags between neural events. The unwrapped coherence phase (symbols), as well as the linear fits from which the delays are extracted (solid lines), are shown in Figure 6A.

With the exception of AC5, all signals have negative slopes, which suggest CN activation preceded the activation in the other areas of interest. A connectivity matrix (Figure 6B) harnessing each ROI as seed against all other ROIs reveals mostly positive delays (red), but the observed delays are not monotonically increasing between CN, SOC, LL, IC, MGB, and AC. Both cortical regions AC4 and AC5 show negative delays (in blue), even up to 5 s, which would mean that their activity precedes that of the other regions. However, a closer inspection reveals that in AC4-5, BOLD contrast was absent (Figure 3 and Figure S2), and thus, where the delays associated with these signals can be considered unreliable. Similarly, MGB5 behaved non-linearly in the low-frequency coherence amplitude, likely due to low BOLD CNR. These unreliable ROIs are thus marked with an asterisk in Figure 6. Taking these exceptions into account, the general trend observed in the ROI-based coherence analysis is an increasing delay with respect to CN1 as the pathway is traversed in the order CN-SOC-LL-IC.

2.4 Voxelwise coherence analysis

While the analysis in Figure 6 is less dependent on exact particulars of neurovascular couplings and hemodynamic response functions, it is still limited by the rather crude choice of ROIs, thus potentially suffering from undesirable partial volume effects. To alleviate this limitation, voxel-by-voxel delay maps were computed from the coherence phase (Figure 7) with the global CN1 time-course used as the seed since its BOLD response was the fastest of the cochlear nucleus ROIs (cf. Figure 3). As observed in the ROI-based coherence analysis, lack of signal may be a source of instability, so the coherence delays were computed only in those voxels exhibiting a coherence magnitude > 0.4.

Most of the pathway yielded positive delays with respect to CN1 (Figure 7), confirming that BOLD responses are elicited earlier in CN1. However, different latencies were observed for different regions, even within the CN1 region. For example, the ventral part of the cochlear nucleus rises faster than its dorsal counterpart (VCN and DCN in zoom 1a). These maps also exhibit interesting latencies between regions: for example, BOLD in the most anterior subregion of the cochlear nucleus (VCA, zoom 2a and 3a in Figure 7), precedes BOLD in the inferior colliculus (zoom 1b,2b,3b) and the lateral lemniscus (zoom 4) and the medial geniculate body in general (zoom 3c). Responses in the medial lemniscus precede those of the ventral lateral lemniscus (ML and VLL in zoom 4), and the central inferior colliculus (CIC) precedes BOLD responses in the external cortex of the inferior colliculus (CIC and ECIC in zoom 2b).

BOLD signals occurring approximately simultaneous with CN1 are depicted in blue and dark red, for example VCA (zooms 1a, 2a, 3a), a contralateral cluster which is located in the sensory root of the trigeminal nerve (s5, zoom 3c) and another small cluster located in the lateral entorhinal cortex (LEnt, panel A, slices 3 and 4). Interestingly, a few negative delays are also apparent, in the dorsal cortex of the inferior colliculus (DCIC in zoom 1b) and some scattered small clusters. Unfortunately, finer segmentations of nuclei would not be as reliable due to the partial volume effect present in these datasets.

The lines in the central diagram shown in Figure 7 represent the neurophysiologically known connections simplified from (Paxinos and Franklin, 2004; Schofield and Cant, 1992). Solid black lines represent connections that can be inferred from the delay maps in Figure 6, while the weak gray lines represent known connections that are not clearly detected. A supplementary video with the temporal reconstruction of these delays is provided online, where the activation, overlaid into a 3D brain template, appears and disappears according to its phase delay in seconds, depicting the BOLD response dynamics in the entire brain.

To evaluate the robustness of the delays map, we computed the full coherence analysis also subject by subject (supplementary Figure S3). The figure shows that subject-by-subject maps do not represent the whole auditory pathway, since many regions are masked out by the coherence amplitude threshold of 0.4. This is even more obvious for those subjects that underwent fewer fMRI runs, but in general the SOC (both ipsilateral and bilateral), ML and LL are not clearly represented anymore. The corresponding GLM-based BOLD maps ($p<0.01$, $t>2$) are plotted in supplementary Figure S4, where the same effect of diminished statistical power is observed, even though both thresholds of 0.4 coherence module and statistical significance t>2 do not necessarily correspond one to one. The lack of statistical power on per-subject analysis in this study precludes a more detailed analysis of intersubject variability, which would require more runs per subject (see Discussion).

## 2.5 Tonotopic maps

Following this general characterization of the auditory pathway, we focus attention on tonotopic mapping. Figure 7 shows t-statistic maps arising from auditory stimulation of a rather broad frequency range, namely, pure tones of either 5, 12, 20 or 35-39 kHz, thresholded for the 3 % highest positive t-values of each frequency at each slice. Some tonotopic organization in the IC is observed, with lower frequencies localizing in the dorsolateral area and higher frequencies corresponding to more ventromedial locations, but an extensive overlap (brown color) is present. Less tonotopy was observed in LL, while signals from the lowest frequencies were detected more in the anterior part of the MGB, and the higher in its more posterior aspect. BOLD signals in CN and SOC also show some degree of tonotopy but again with some overlap.

Supplementary Figure S5 shows the subject-level global tonotopy analysis, again revealing that tonotopy was less clearly observable in each individual subject, likely due to the lower statistical power in the subject level. The low frequency tones produced a more dorsal response in the IC clearly and consistently across subjects, and elicited a stronger activation in other areas such as MGB, LL and SOC. It is more obvious in subject 3, which had 6 fMRI runs (2 per frequency), again suggesting the importance of the statistical power. For this specific subject, some degree of tonotopy was also present in the SOC and LL.

Finally, we aimed to study, through Experiment 2, whether fMRI could be used to map 1 kHz-separation, corresponding to small spatial correlations within small frequency differences. A strong overlap was observed between these very close frequencies on both group- and subject level analyses (Figure S6). In the case of the 1 kHz-separation additional study, little distinct localization was observed for the different frequencies at the subject level. Most frequencies overlap in the IC, and again it is clear that the higher the amount of data acquired (increasing n), the higher the significance in t-values in the auditory pathway, and a slightly better sepration of tones can be observed.

# 3  Discussion

The auditory pathway is a highly distributed network, which exhibits extensive plasticity in development, adulthood, as well as in disease (Bandyopadhyay et al., 2010; Polley et al., 2008; Schreiner and Polley, 2014; Yu et al., 2007). The mouse auditory pathway, in particular, is an attractive target for imaging, given that mice – as compared to rats – may be more amenable to transgensis, optogenetics, and animal models of disease. Manganese enhanced MRI (MEMRI) (Yu et al., 2008, 2007, 2005) was previously used to map the mouse auditory pathway using very long stimulation periods that elicited preferential $Mn^{2+}$ uptake, which, in turn, provided contrast in parts of the pathway. Although these MEMRI-based studies enabled a characterization of the auditory system and to some extent, its plasticity, such studies are different than BOLD studies in that the stimulation is performed offline and thus the images represent a "static" view of accumulating effects. MEMRI studies may also be somewhat restricted by the need to stimulate animals for many hours, which may induce stress, as well as by issues with $Mn^{2+}$ toxicity that may limit longitudinal measurements in the same mouse. On the other hand, BOLD fMRI may offer a complementary view of the pathway: it enables the detection of BOLD responses upon event- or block-paradigms, and BOLD response latencies can be surrogate markers for activation dynamics. Furthermore, BOLD fMRI lends itself nicely to longitudinal studies.

Thus, the main objectives of this study were: (1) to develop an experimental setup enabling auditory mapping in the *in-vivo* mouse using fMRI; (2) to establish whether the auditory pathway-related regions (e.g., (Malmierca and Ryugo, 2012)) would exhibit significant BOLD responses; (3) to characterize the ensuing BOLD signals; (4) to study the BOLD activation latencies; and (5) to test whether the pathway's critical features, such as distinct latencies for each auditory center signals or tonotopy, could be mapped *in vivo* in the mouse.

The experimental setup presented here indeed offered auditory pathway mapping in the mouse using fMRI. We observed highly significant BOLD responses along most of the auditory pathway, and in fact, they could be clearly discerned with the naked eye in relevant ROIs even for a single run in a single animal (Figure 2). These BOLD responses were typically < 1 % of the baseline signal, but the datasets were of the quality that clearly enabled a full characterization of the signals. Our use of the cryoprobe and its > 2 signal to noise enhancement was quite central to this high detectability.

Regions exhibiting highly significant activation upon pure tone stimulation in both GLM fitting and coherence analysis included the CN, SOC, LL, IC, and MGB – all areas known to be prominently involved in the auditory pathway (Malmierca and Ryugo, 2012). The 5 kHz experiment (cf. Figure 5) was characterized by the strongest activation and additionally to the regions mentioned above, activation was noted also in the crus 1 of the ansiform lobule of the cerebellum (Crus1), and bilaterally in SOC. Activity in Crus1 had already been reported in auditory studies in both rats (Lau et al., 2013) and humans (Petacchi et al., 2005). The bilateral SOC activation was observed mainly for the frequency with the highest SPL of all stimuli prescribed in this study (5 kHz), and such high SPLs are known to elicit activity from the descending auditory pathway (olivocochlear circuit) on the ipsilateral SOC (Zhang et al., 2013). The higher harmonic distortion of this frequency in our stimulation system may have activated neurons in low and high frequency regions of the ipsilateral SOC, that did not necessarily activate when a cleaner sound was presented via the other frequencies (Watson et al., 2012). For the 35-39 kHz experiment, bilateral SOC activation was also observed, though not as strongly as for 5 kHz (Figure 5).

Another feature of the BOLD responses observed in this study was the frequency selectivity that enabled observation of tonotopy (Figure 7) in most of the auditory pathway when sufficient statistical power was available. The choice of t-map threshold at the highest 3% positive t-values reflects a compromise between spatial specificity and sensitivity: since the BOLD mechanism is also less spatially specific than the underlying neural activity, it could be assumed that the most highly significant voxels indeed correspond to the areas that are most strongly associated with a given frequency. This was clearly the case, as the IC exhibited the expected dorsolateral to medioventral partition of lower to higher frequencies, respectively.

Activity in one crucial area along the auditory pathway – the AC – could not be detected in this study, neither in ROI-based analysis or in GLM or coherence analyses. Several factors may have contributed to this observation. First, BOLD responses may be smaller in AC due to local properties of the vasculature distribution or neurovascular coupling. In such a case, a $T_2/T_1$ sequence such as the FISP used in this study may not have sufficient contrast to noise to detect the underlying BOLD. Another potential sequence-related culprit is the relatively high resolution used here, which can diminish BOLD responses due to the smaller susceptibility distribution within smaller voxels. Furthermore, it has been shown in rat that chronic exposure to high SPL's can suppress AC response (Lau et al., 2015). Although the sound exposure in this study was not long lasting, it had a SPL that was well above the one used in the rat study (85 dB), which could potentially suppress AC activity. It also worth mentioning that anesthesia/sedation can have a suppressing effect on the higher-level structures of the auditory pathway (Cheung et al. 2011), which may have made detection of AC activity even more difficult in this study. Future studies harnessing gradient-echo EPI to increase the susceptibility weighting as well as stimulation during quiet periods in the scan to decrease AC activity suppression, may tackle these issues and reveal BOLD in the mouse AC.

In recent years, many BOLD features have been reported *in-vivo* in the rat auditory pathway (Cheung et al., 2012a, 2012b, Gao et al., 2015a, 2015b, 2014, Lau et al., 2013, 2011b). It can thus be interesting to compare the BOLD responses between the species (noting of course differences in experimental parameters and anesthesia between studies).

> **Auditory cortex**. Auditory experiments in the rat revealed activity in the AC (Cheung et al., 2012a, Zhang et al., 2013), contrary to the present findings in the mouse, where no BOLD was detected. In the rat, the amplitude of AC activation was always the smallest of the auditory pathway structures, again suggesting that stronger susceptibility weighting may be required to amplify the BOLD responses in the mouse AC.
>
> **Other auditory pathway regions**. BOLD patterns were observed in the rat CN, SOC, LL, IC and MGB, consistent with our findings. In our study, CN, IC, and SOC evidenced increases in signal of 0.5% during activation and LL and MGB showed somewhat smaller increases (around 0.2-0.4%). Although the absolute BOLD magnitudes are different (n.b., different pulse sequences were used), these findings are consistent with the rat studies (Cheung et al., 2012a), where the CN, IC and SOC exhibited stronger activation than LL and MG. In addition, we observed that lower frequency stimuli produced more robust BOLD responses in the mouse (Figure 5), consistent with the earlier studies performed in rats (Cheung et al., 2012a, Cheung et al., 2012b).

**BOLD signal shape and latency.** Cheung et al showed that the time-to peak for all auditory pathway structures in rats was around 10 s (Cheung et al., 2012a). In another study (Lau et al., 2015) harnessing higher temporal resolution (1 s), the time-to-peak in IC was around 3 s. In our study, most structures involved in the auditory pathway peaked faster, around 8 s (Figure 3 and Figure S2). These species-related BOLD time-to-peak differences may be due to neurovascular couplings or due to specifics of the experimental design and pulse sequences used in the different studies. Interestingly, in the IC activation plots (Figure 3) an overshoot followed by a plateau was observed in this study, consistent with findings in rats stimulated at similar frequencies (Cheung et al., 2012a). This may reflect similar sensitivity of mouse and rats to frequencies in this specific range (Heffner et al., 2001). Additionally, the CN exhibited a second peak in activation when the stimulation was turned off, consistent with Cheung et al's results in the rat (Cheung et al., 2012a). This may be due to an ON-OFF response pattern mediated by the auditory system, which triggers neuronal firing when the stimulus is turned off (Henry, 1985).

**Tonotopy.** Our results in the mouse indicated clear tonotopy in several areas along the pathway, consistent with previous studies in the rat (Cheung et al., 2012a, 2012b). However, unlike the rat studies, we additionally observe quite extensive frequency overlaps in IC, ML and LL. The sources for these overlaps may include: (1) lack of statistical power, mainly due to the medetomidine sedation, which does not produce a deep sedation in mice (Meyer, 2008) and prevents the acquisition of more runs per subject; (2) potentially inherent differences in rat vs. mouse haemodynamic responses, which may also affect the spatial specificity of the signals; (3) the block paradigm stimulation used in this study, which is not necessarily ideal. A more sophisticated design like the swept source technique previously described in rats (Cheung et al., 2012b) may provide a clearer tonotopic segmentation; (4) the mouse measurements may have been more susceptible to partial volume effects in the slice direction due to our slice thickness, and (5) our thresholds were chosen with the same criterion per frequency whereas in other reports a less stringent thresholding was chosen which may have masked potential overlaps (e.g., Figure 5 in (Cheung et al., 2012a)).

All these suggest that there are both similarities and differences in the mouse and rat auditory pathway; the areas activated are similar and some patterns are similar, but time-to-peaks and specific BOLD dynamics may be different between species.

The data analysis in the present study included a comparison of data-driven coherence analysis (which was also previously also used in a rat study of the auditory pathway (Cheung et al., 2012b)) and model-based GLM fitting. Both approaches provided very similar maps (Figure 4 and Figure 5). Since the former method is to first order independent of the specific input model for the HRF, the similarity between GLM and coherence maps suggests that hemodynamic responses may not be very different along the pathway. Coherence analysis also provides an added benefit, in that while its magnitude is analogous to the GLM's t-statistics, it also reports the coherence phase, which represents delays or latencies in BOLD responses. The ROI analysis clearly showed the expected linear behavior in the low frequency region (Figure 6), from which the delays can be easily computed both ROI-wise and voxelwise (Figure 7). In general, shorter delays were found for CN and SOC and longer for IC and LL, which is in agreement with the system pathway synaptic order (Malmierca and Ryugo, 2012). In addition, the delay maps suggest a heterogeneity in many regions, that can reflect either differences in neurovascular couplings, or real differences in underlying neural activity. For example, in the IC, CIC and ECIC exhibit spatially distinct phases (red and yellow colors, respectively, in Figure 7F) which may suggest that they are involved in different phases of the sound processing in the IC. Nevertheless, these delay maps should be viewed with caution in the context of the underlying neural activity, since BOLD responses characterized here likely do not clearly distinguish the detailed ascending/descending dynamics, which are known to be extremely rapid (Heil, 2004). Future electrophysiological assessment would be required to establish the correspondence of the BOLD latencies with the underlying neuronal activity.

    Another aspect of the analysis in this study was the statistical power, whose necessity was clearly demonstrated. While group-level effects were rather clear, the activation maps of individual animal showed lower t-statistics and much higher variability. In addition, the fewer runs per subject, the poorer the activation maps that were observed in both coherence and GLM analyses (Figure S3 and Figure S4). The main reason for the uneven sampling of BOLD time series in different animals was that the animals started moving and slowly withdrawing from medetomidine at quite variable times. Hence, in some animals that responded well to the sedation, up to n = 6 runs could be accomplished while in others, much fewer runs were completed before the experiment was aborted. This made a direct comparison of inter-subject variability much more difficult in this study. In the future, we propose to utilize etomidate, an anesthetic that allows for very long and stable sedation (Heil, 2004) or using awake animals (M. Desai et al., 2011). Recent experiments in our group show that this is indeed a better way of performing such experiments and reaching much better results on a per-subject level.

Several limitations of this study can be identified. First, perhaps quite general to nearly any fMRI-based study, is the indirect nature of the BOLD mechanism vis-a-vis the underlying neural activity, which makes it very difficult to assess the direct circuit dynamics, in terms of, e.g., inhibitory, excitatory, and modulatory characteristics. Our results are thus more limited to reporting the locations of activity, and the latencies of the more distinct early, intermediate, and late phases discerned above. Future experiments with either line-scanning techniques (Yu et al., 2013) and/or using nonBOLD fMRI (Le Bihan et al., 2006), along with event-related paradigms (Schlegel et al., 2015b), could be used to access much faster dynamics and to achieve a better spatial specificity in the tonotopic maps. Moreover, stimulating with logarithmically distributed frequencies, such as octaves or fractions of octaves (Barnstedt et al., 2015), or perhaps with some custom acquisition paradigm like the swept source proposed by Cheung et al. (Cheung et al., 2012b) may allow a better visual assessment of the spatial organization in the auditory system. Additionally, although our resolution of 150 x 150 x 650 $\mu m^3$ can be considered very high for mouse fMRI, it still clearly suffers from partial volume effects which may prevent the evidence of finer tonotopies. The different LL nuclei, for example, are not completely resolved in our experiment. Better resolution can be achieved by compromising the number of slices in favor of averaging higher resolution images at the same temporal resolution. On the other hand, the robustness of the coherence analysis to the delineation of the seed ROI should be further evaluated, for example by developing a criterion for an automatic definition of the seed region. Finally, another limitation is the medetomidine sedation. Training may allow for such experiments to be conducted in awake mice (M Desai et al., 2011; Harris et al., 2015), which could be beneficial for studying global brain responses upon the delivery of more natural stimulus, such as, e.g., pup vocalizations. Unfortunately, due to the unbalanced distribution of males/females in our study, we could not assess gender differences with the different frequency cues, but such comparisons are definitely interesting future lines of research in combination with natural sounds.

In conclusion, the mouse auditory was mapped using high resolution BOLD fMRI. Highly significant BOLD responses were observed, delineating the entire pathway except for the cortex, and the GLM results were in agreement with the data-driven coherence analysis. This lays the foundations for future studies harnessing the mouse's full transgenic potential for studying the longitudinal aspects of the pathway, including plasticity and aberrations in disease, as well as specific circuit modulations using optogenetics.

# 5 Figures/Tables

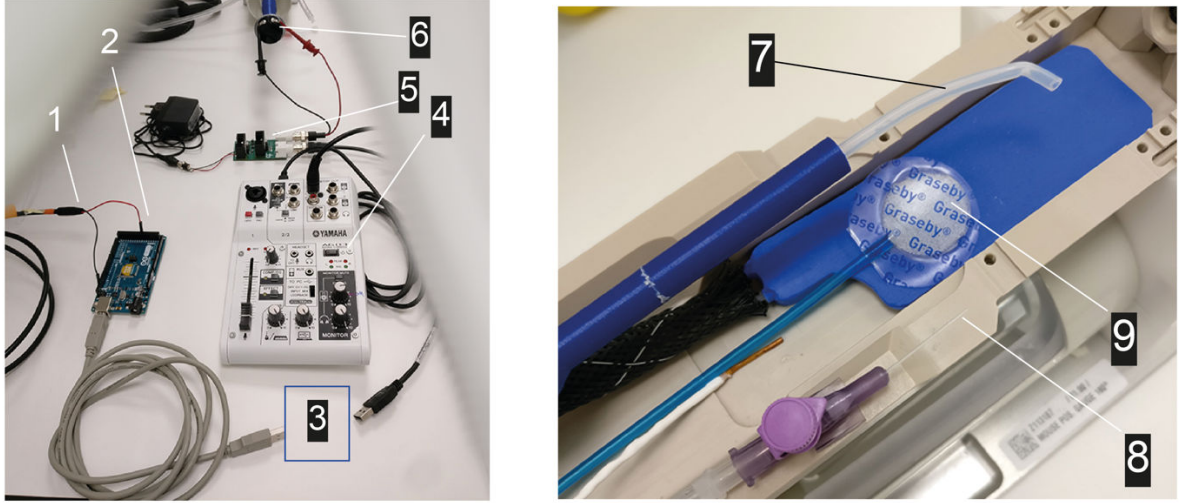

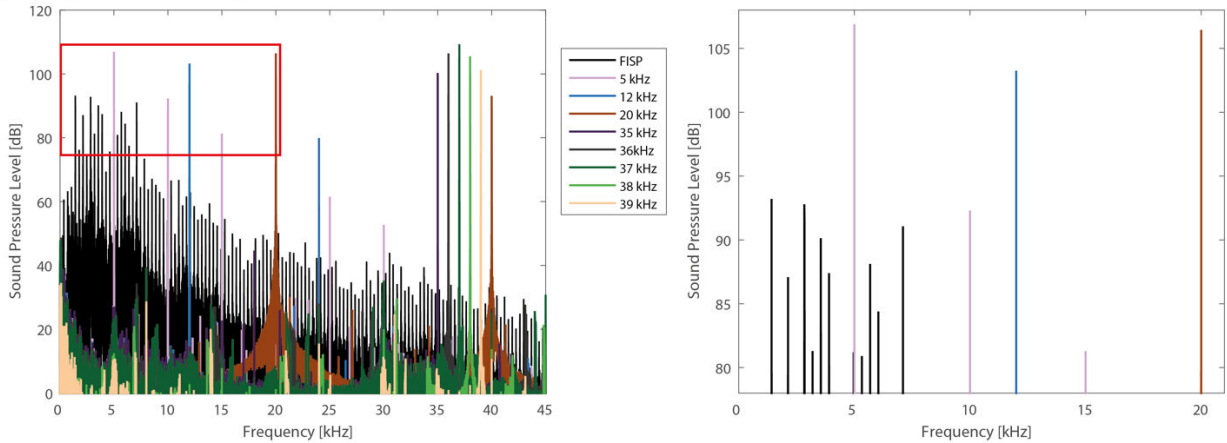

**Figure 1.** Experimental setup for auditory pathway mapping in the mouse. **(A)** The different elements essential for the delivery of the stimuli. (1) TTL trigger signal from the scanner, (2) Arduino, (3) auxiliary computer which generates the sounds, (4) sound board, (5) amplifier, (6) speaker -docked to the main nylon tube-, (7) Tygon tip for ear insertion, (8) medetomidine catheter, (9) respiratory sensor. **(B)** A spectral characterization of the pure tones delivered in this study, along with the intrinsic noise arising from the specific pulse sequence used for acquiring

the functional maps. Notice that the pure tones are always at least 6 dB higher compared to the scanner noise.

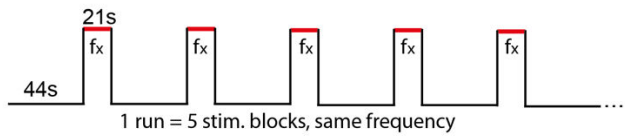
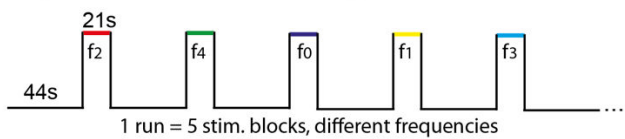
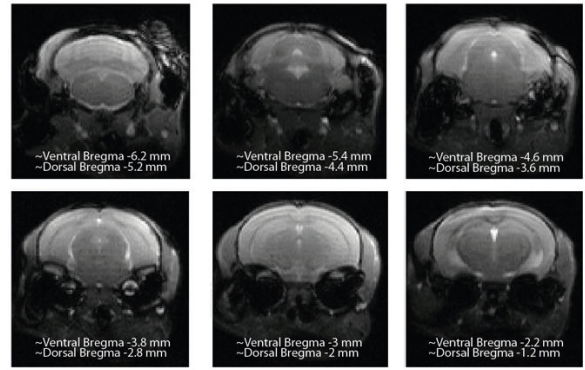
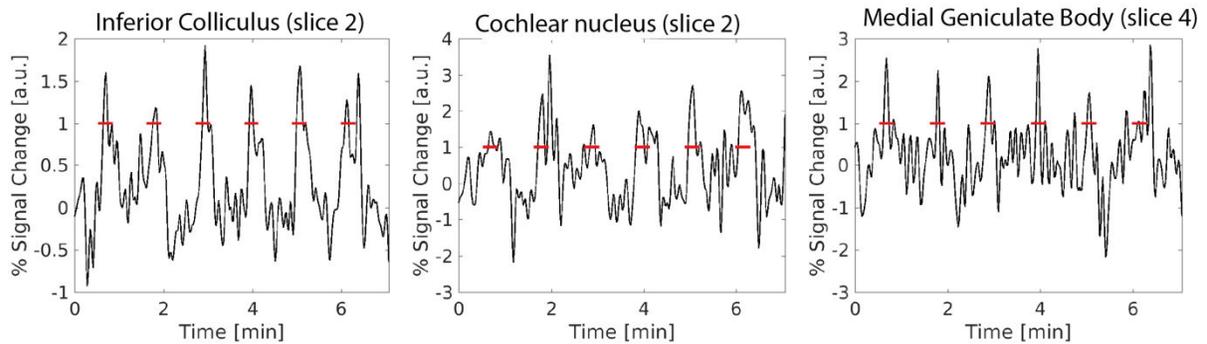

**Figure 2.** Paradigms and robustness of raw data. **(A)** The paradigm used for the global tonotopy (experiment 1) and the 1kHz tonotopy (experiment 2). Notice that in experiment #1, a single tone with frequency $f_x$ was repeated, while in experiment #2, the tones were randomized. **(B)** The raw data arising from a typical fMRI experiment. Using the cryoprobe, these raw images present excellent SNR even at the high spatial and temporal resolution they were acquired with. Localizations with respect to Bregma should be considered approximate since the slices were not exactly axial and the slice separation is 0.8mm. **(C)** Likewise, ROIs placed in prominent regions along the pathway in a single mouse and a single run reveal the BOLD responses even with the naked eye (a typical dataset is shown from one single representative mouse). The ROIs delineation can be found in Figure S1. The red bar indicates stimulation with a 5 kHz pure tone.

# BOLD timeseries in selected ROIs

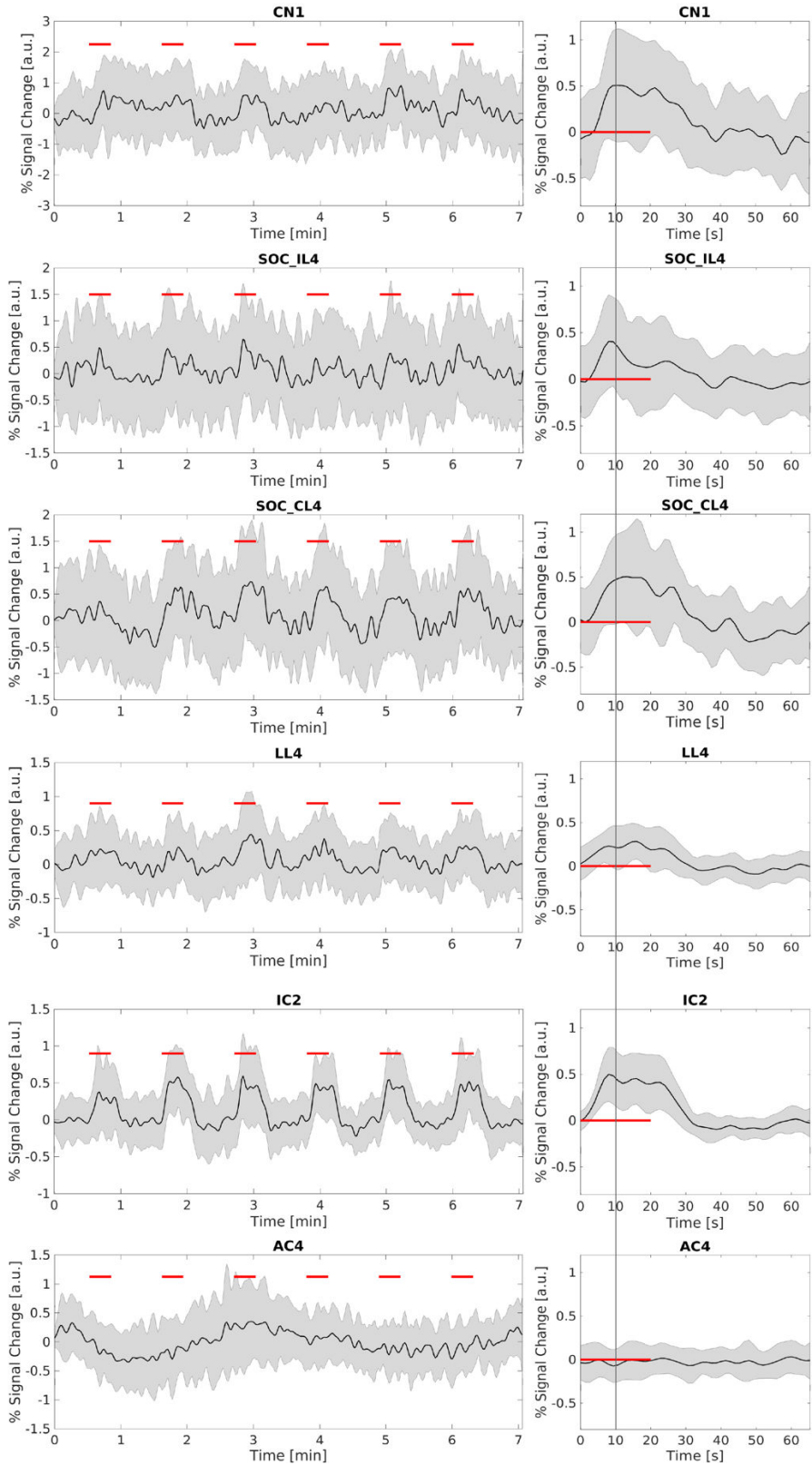

(A) corrected timeseries  (B) average cycle

**Figure 3.** BOLD responses in representative auditory pathway ROIs for the pool of all experiments. (A) The raw time series for all animals, n = 44 runs (registered to the same space), including the mean (solid line) and standard deviation (shaded regions). (B) The averaged cycle (solid line) and standard deviation of the averaged cycle (shaded plots) with a vertical line to guide the eye at t=10s from the stimulation start. BOLD responses were easily observed in all ROIs following stimulation (red bar) except for the auditory cortex. Six ROI representatives are shown in this figure, the remaining ROIs are shown in Figure S2 and the ROI delineations are depicted in Figure S1 [CN1=cochlear nucleus in slice 1, SOC_IL4=ipsilateral superior olivary complex in slice 4, SOC_CL4=contralateral superior olivary complex in slice 4, LL4=lateral lemniscus in slice 4, IC2= inferior colliculus in slice 2, AC4=auditory cortex in slice 4].

## A) GLM *t-value* maps

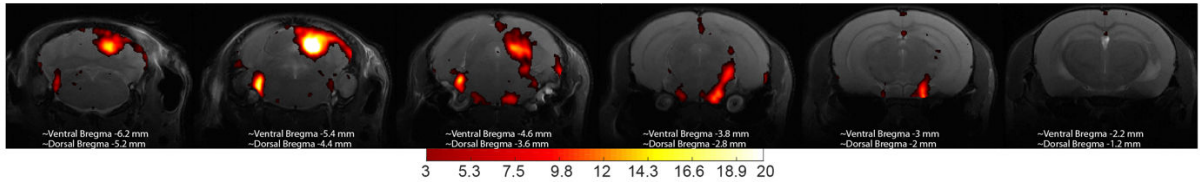

## B) Coherence (to CN1) amplitude maps

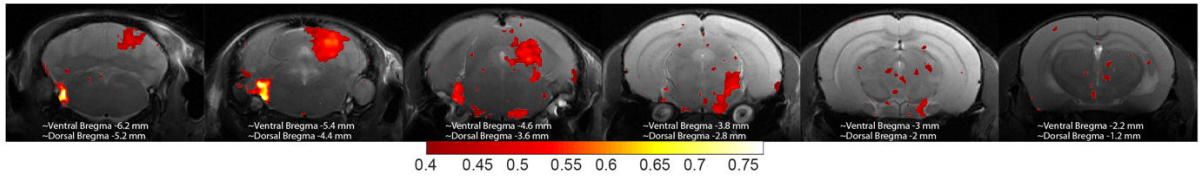

**Figure 4.** Representative t-statistic (panel A) and coherence maps (panel B) of the global auditory response from the pool of all experiments (n=44). Notice the very highly significant t-statistics, representing good detection of activated voxels in the auditory pathway, and the similar regions exhibiting high coherence with CN1. No cluster size thresholds applied. Localizations with respect to Bregma should be considered approximate since the slices were not exactly axial and the slice separation is 0.8mm.

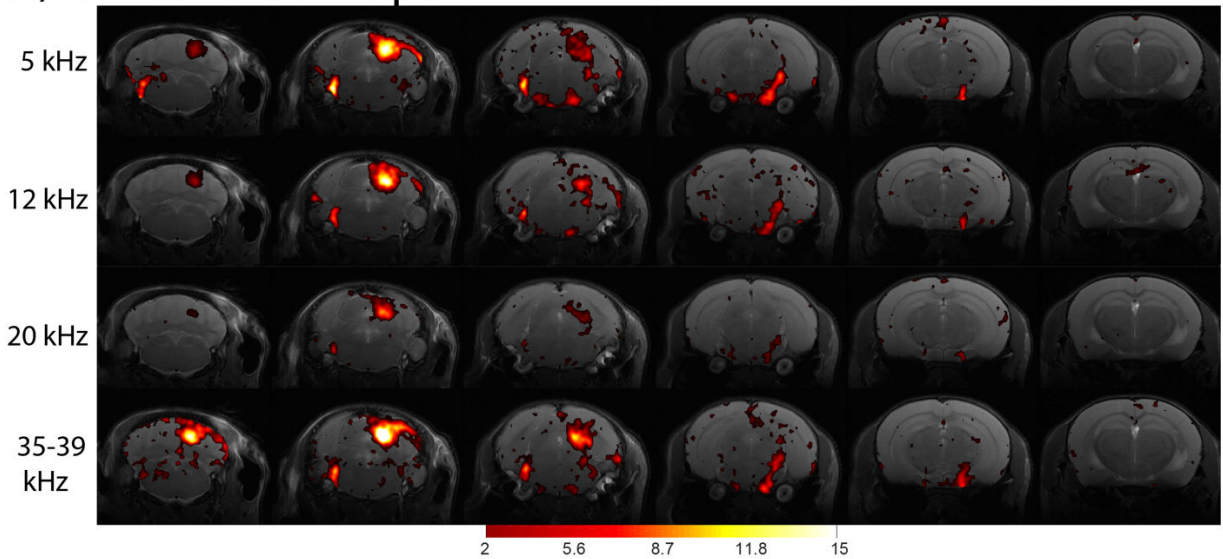

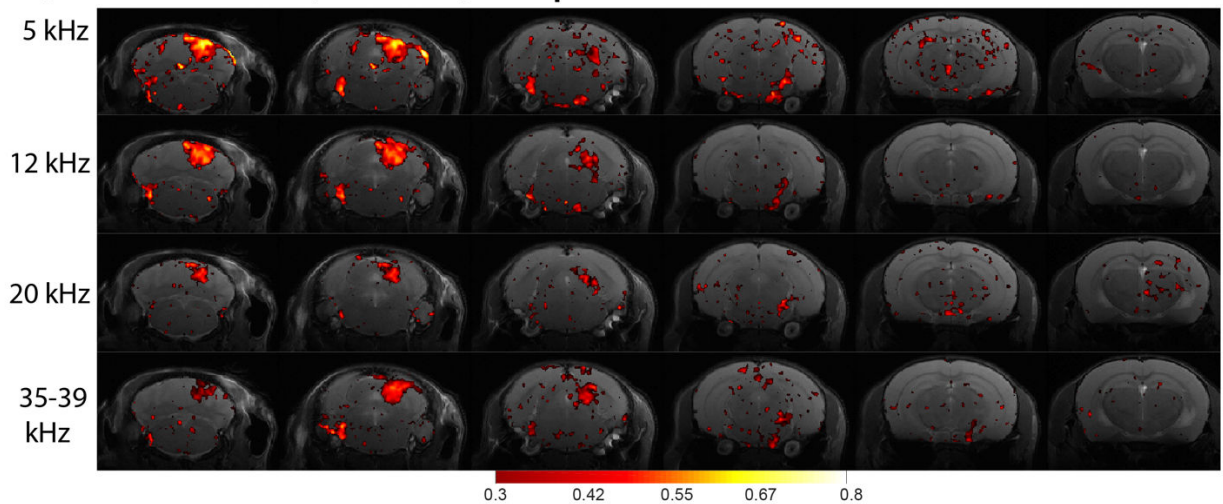

**Figure 5.** A) BOLD representations for [from up-down]: all stimulation frequencies analyzed independently, 5 kHz, 12 kHz, 20 kHz and the joint analysis of 35-39kHz experiments. The fMRI runs distribution is described in Table1 (n=6 for 5, 12, and 20kHz, n=26 for 35-39kHz. B) Coherence (to CN1) amplitude maps for the same stimulation frequencies. The responses are robust along the auditory pathway, although the 20 kHz tone provides slightly less activation. T-values over 8 are found in the ipsilateral CN, contralateral IC, contralateral SOC, LL, and MGB.

Similarly, the coherence amplitudes show high values in the mentioned regions. No cluster size thresholds have been applied.

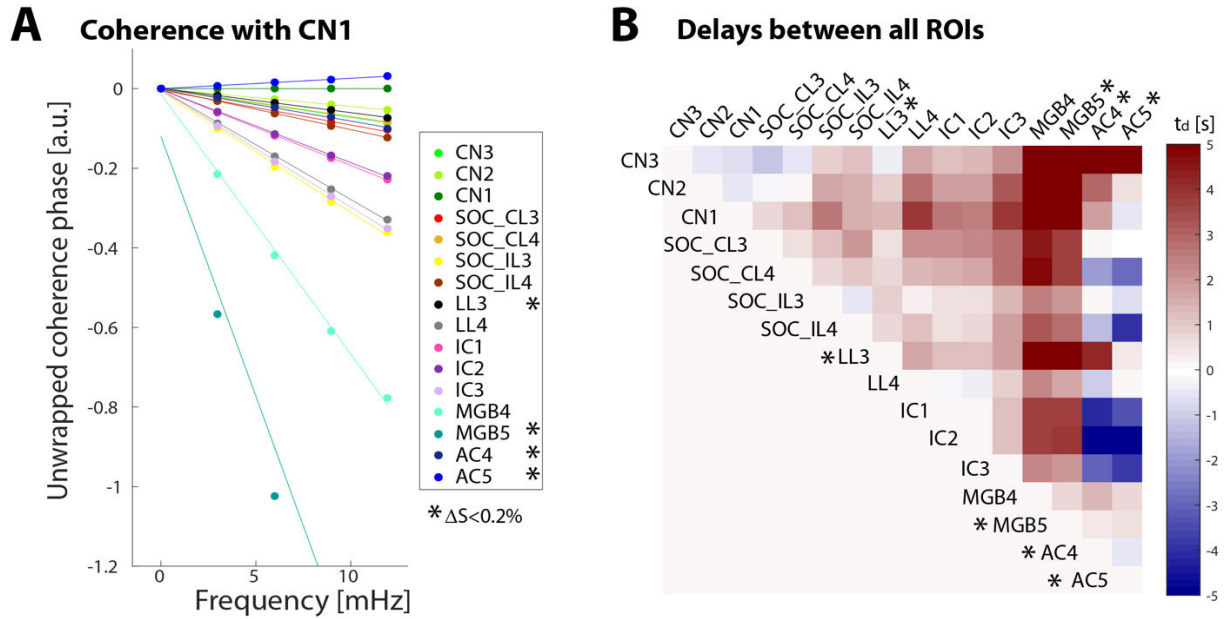

**Figure 6**. Coherence analysis for the ROIs extracted from the average dataset (n = 44 runs). (A) The coherence phase in the lower frequencies showed a clear linear response, as expected, from which temporal delays could be extracted. (B) The ensuing delay (connectivity) matrix, utilizing each ROI as the seed against all other ROIs. The color scale reflects td, the delay in seconds with respect to the seed region. *regions were the BOLD signal was below 0.2%.

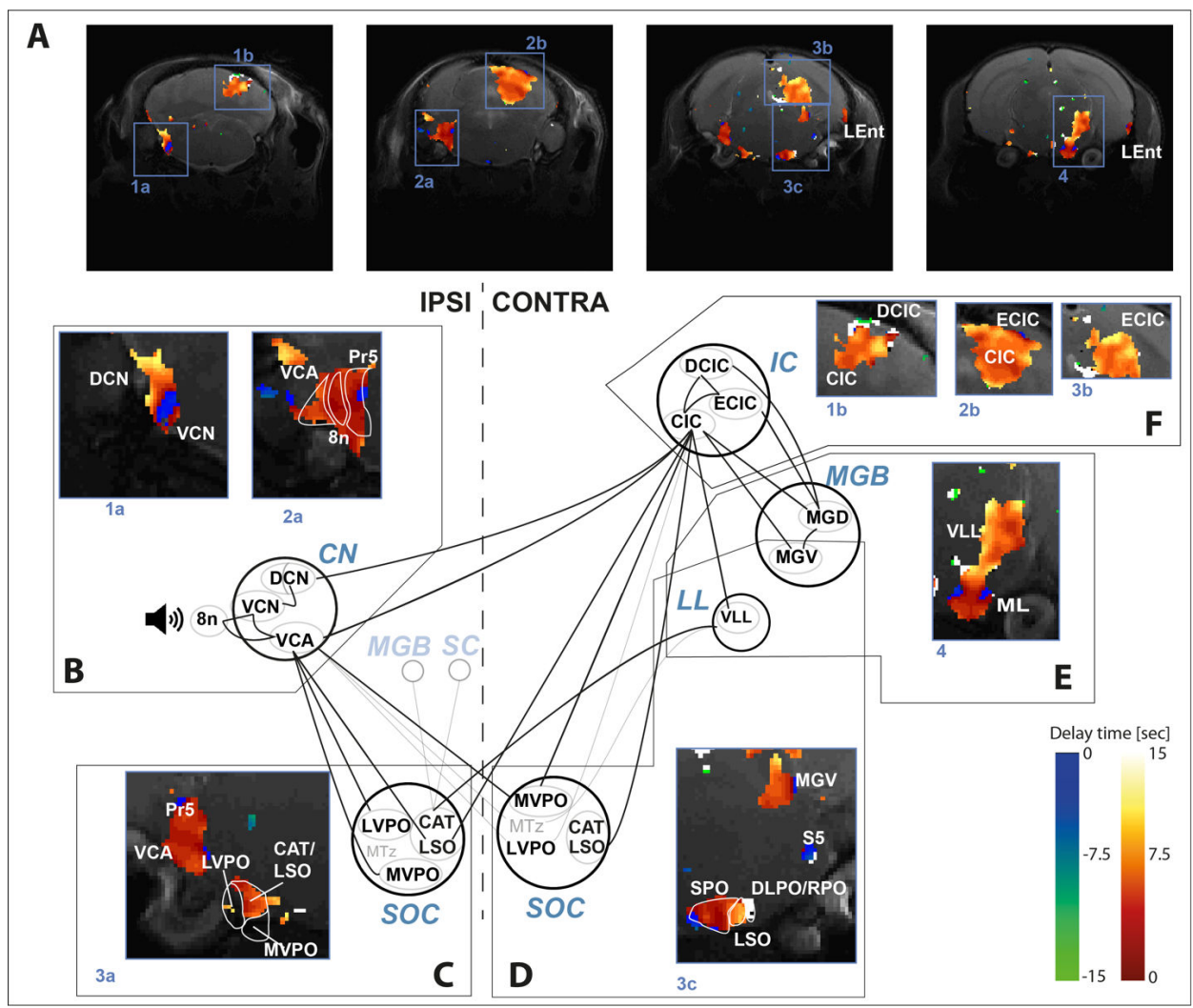

**Figure 7**. (A) Voxel-by-voxel delay maps reveal the latencies of the auditory pathway (computed from the average dataset, n = 44 runs). (B) Detail from the cochlear nucleus, (C) Detail from the ipsilateral supraolivary complex and the most anterior part of the cochlear nucleus, (D) Detail from the contralateral supraolivary complex and medial geniculate body (E) Detail from the medial lemniscus and lateral lemniscus, (F) Detail from the inferior colliculus. Since the more posterior CN ROI was used (CN1), most of the delays are positive (hot scale), showing distinct temporal latencies for several regions and even within each region. The schematic in the center was simplified from *(Paxinos and Franklin, 2004; Schofield and Cant, 1992)*.

Global tonotopy

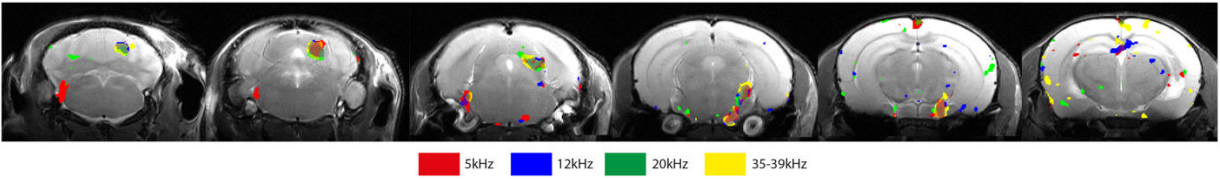

**Figure 8**. BOLD tonotopy in the mouse brain upon stimulation with widely different frequencies (n = 12 animals, n = 44 runs). The IC, SOC, LL and MGB shows some degree of tonotopy, but they all exhibit an extensive overlap.

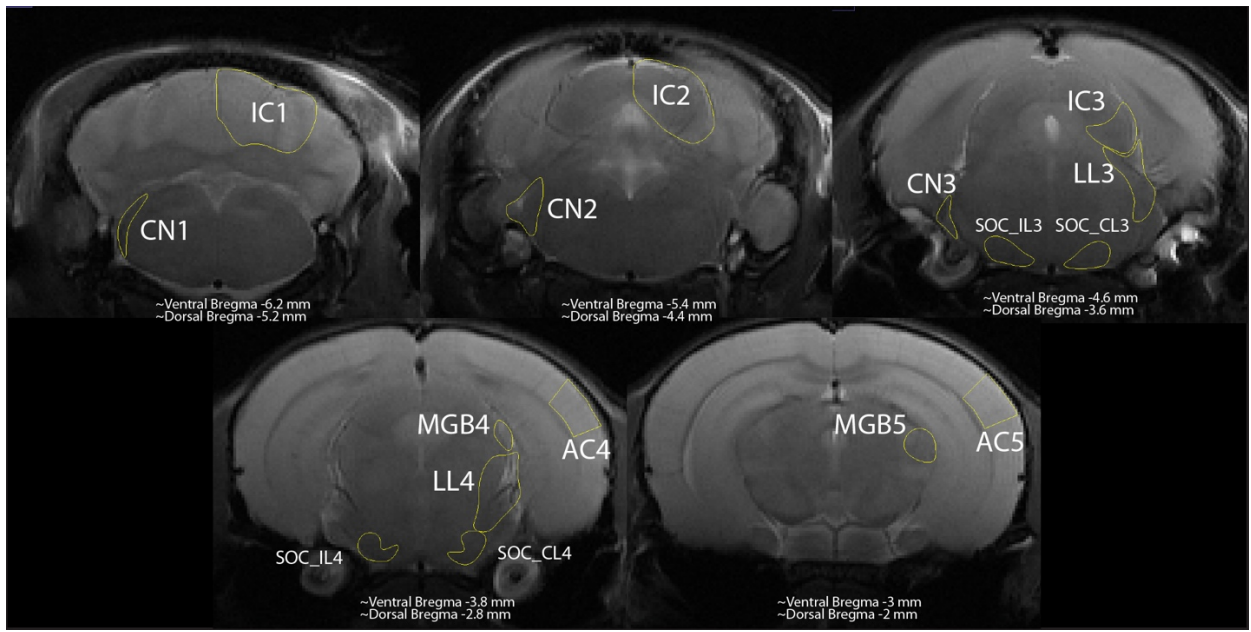

**Figure S1**. Delineation of the regions of interest along the auditory pathway. The delineation was performed with ImageJ and according to Paxinos and Franklin atlas (Paxinos and Franklin, 2004). Localizations with respect to Bregma should be considered approximate since the slices were not exactly axial and the slice separation is 0.8mm. Those regions which comprised more than one slice were stored independently and were not averaged, to provide a more detailed exploration of the auditory pathway, and they were tagged with their formal abbreviation followed by the number of their slice. To distinguish between ipsilateral and contralateral superior olivary complexes they were labeled as "SOC_IL" and "SOC_CL" respectively. Full list of regions: CN=cochlear nucleus, IC=inferior colliculus, LL=lateral lemniscus, SOC_IL=ipsilateral superior olivary complex, SOC_CL=contralateral superior olivary complex, MGB=medial geniculate body, AC=auditory cortex.

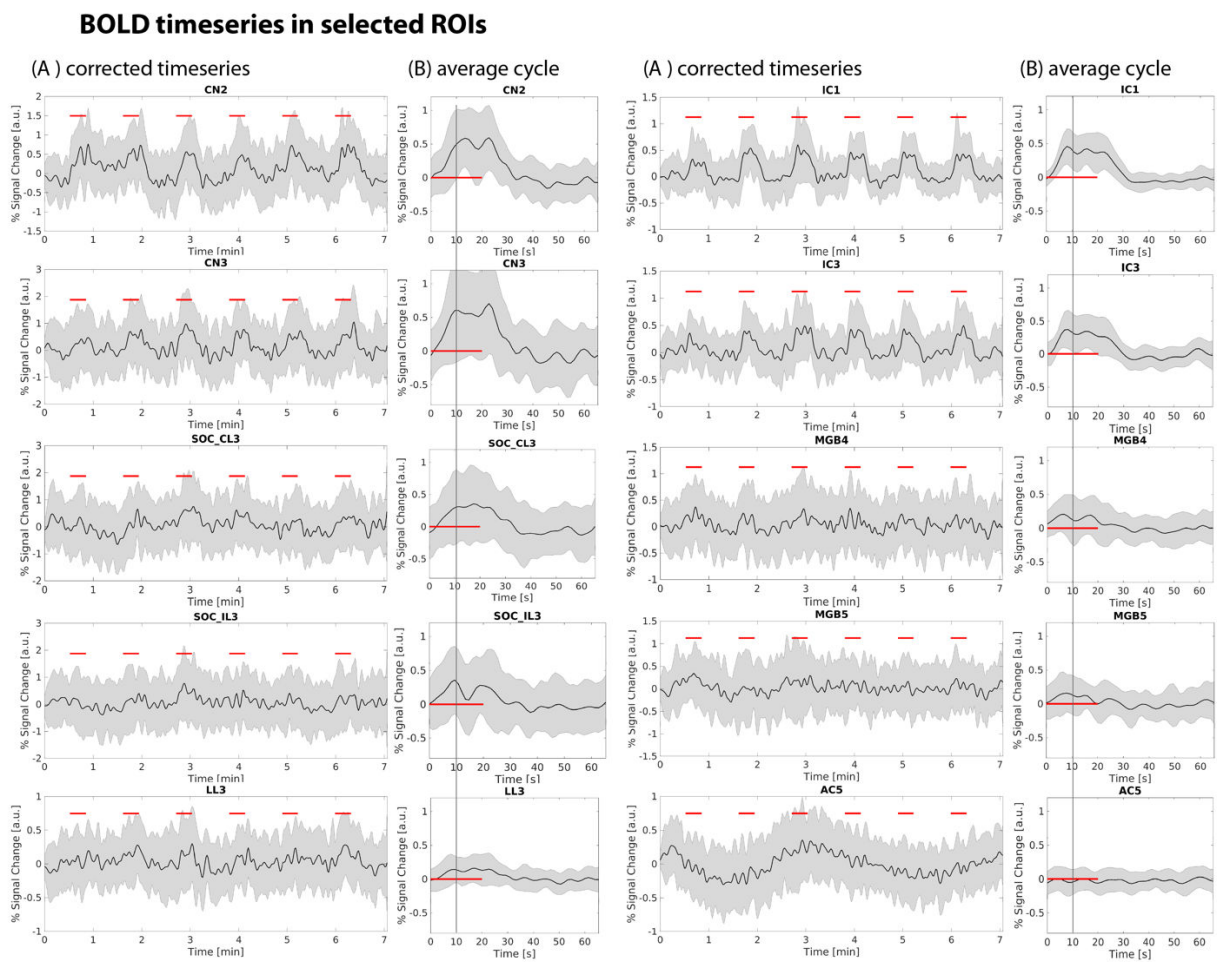

**Figure S2**. BOLD responses in auditory pathway ROIs for the pool of all experiments. **(A)** The raw time series for all animals, n = 44 runs (registered to the same space), including the mean (solid line) and standard deviation (shaded regions). **(B)** The averaged cycle (solid line) and standard

deviation of the averaged cycle (shaded plots) with a vertical line to guide the eye at t=10s from the stimulation start. BOLD responses were easily observed in all ROIs following stimulation (red bar) except for the medial geniculate body and the auditory cortex. The ROI delineations are depicted in Figure S1 [CN2=cochlear nucleus in slice 2, CN3=cochlear nucleus in slice 3, SOC_IL3=ipsilateral superior olivary complex in slice 3, SOC_CL3=contralateral superior olivary complex in slice 3, LL3=lateral lemniscus in slice 3, IC1= inferior colliculus in slice 1, IC3= inferior colliculus in slice 3, MGB4=medial geniculate body in slice 4, MGB5=medial geniculate body in slice 5, AC5=auditory cortex in slice 5].

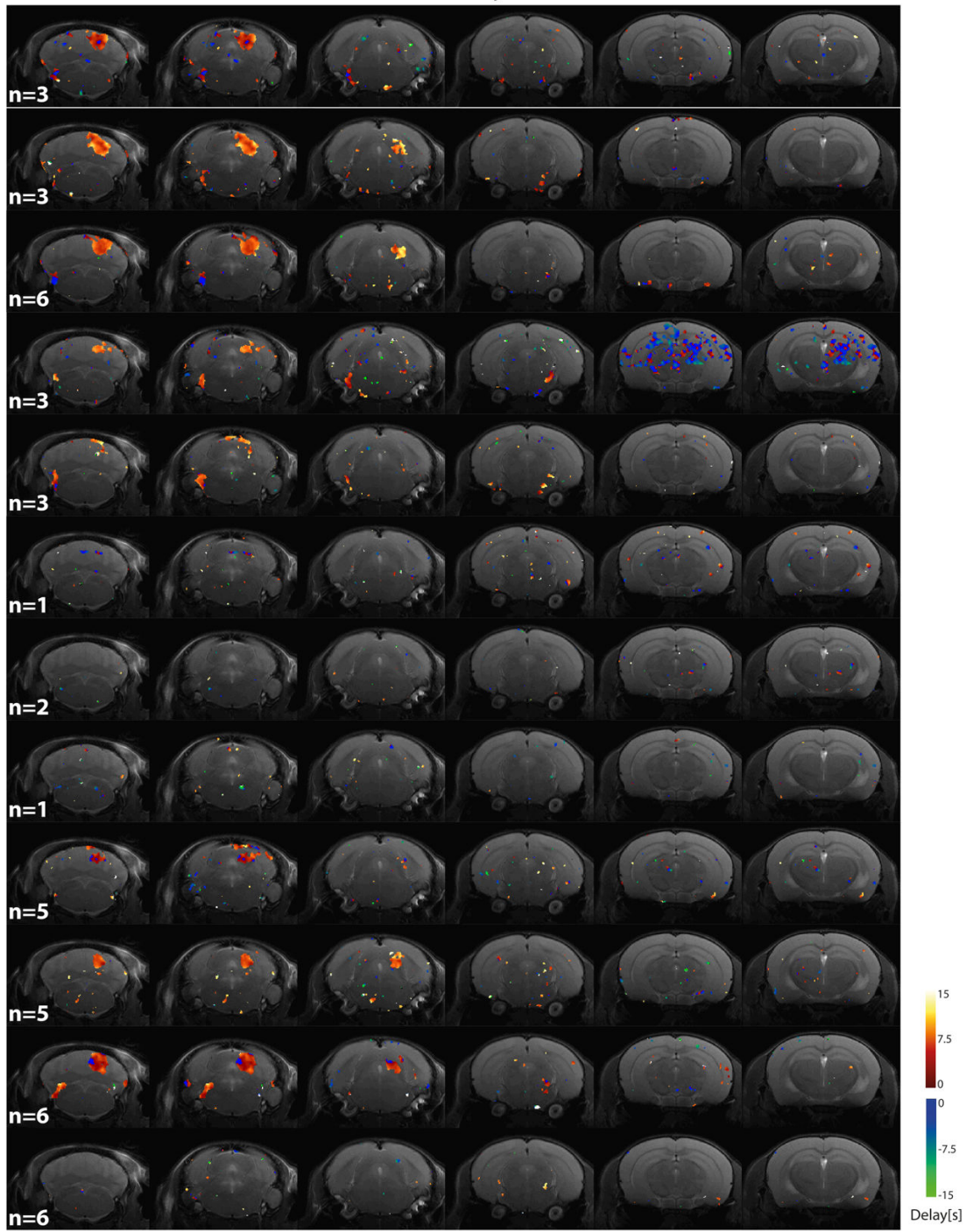

**Figure S3.** Voxel-by-voxel coherence delay maps per subject. Most of the pathway regions did not exceed the coherence module of 0.4, proof of the signal loss suffered due to the splitting of the data. Their corresponding GLM maps are shown in supplementary Figure S4.

## GLM Maps

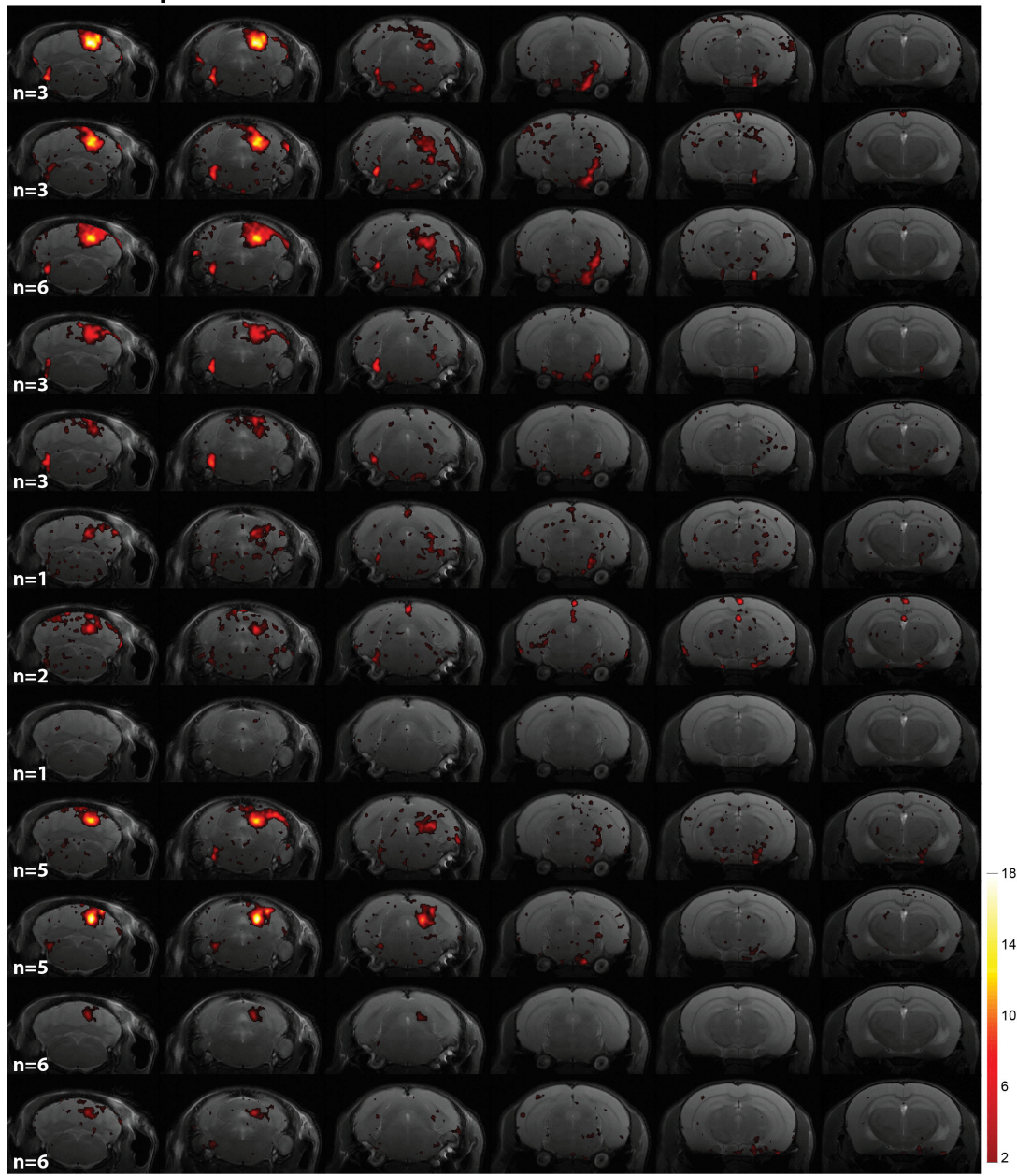

**Figure S4.** GLM BOLD maps per subject. The significance of the auditory regions decreases as compared to the map of the pool of all subjects represented in Figure 4. The strength of the BOLD signal, or additionally the coherence amplitude, could be used as an exclusion criterium for performing coherence delay analysis. Note: for convenience, they were all plotted on top of the reference anatomical RARE which was used as the reference for all brains normalization.

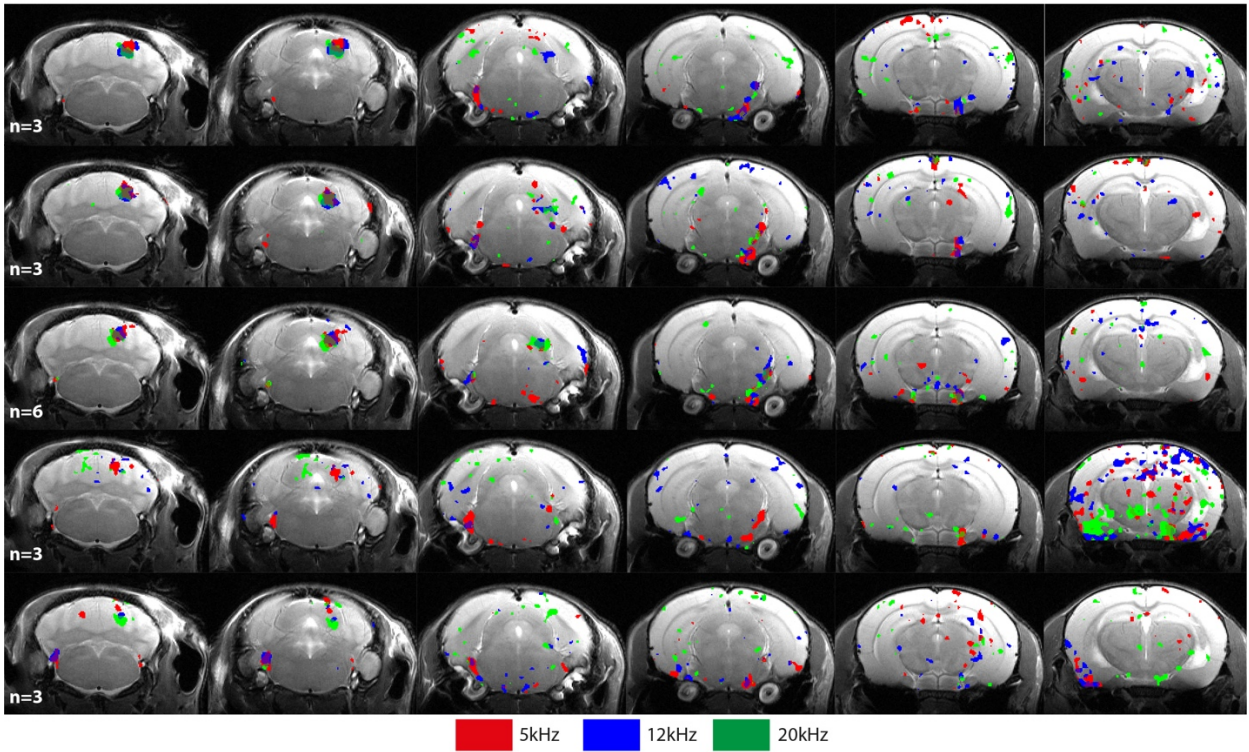

**Figure S5**. Individual tonotopic maps per subject from experiment 1 built from the top 3% positive t-values. Some degree of tonotopy is observed mainly in the IC, more clearly in subject 3 which had 2 runs per frequency and therefore exhibits higher statistical power

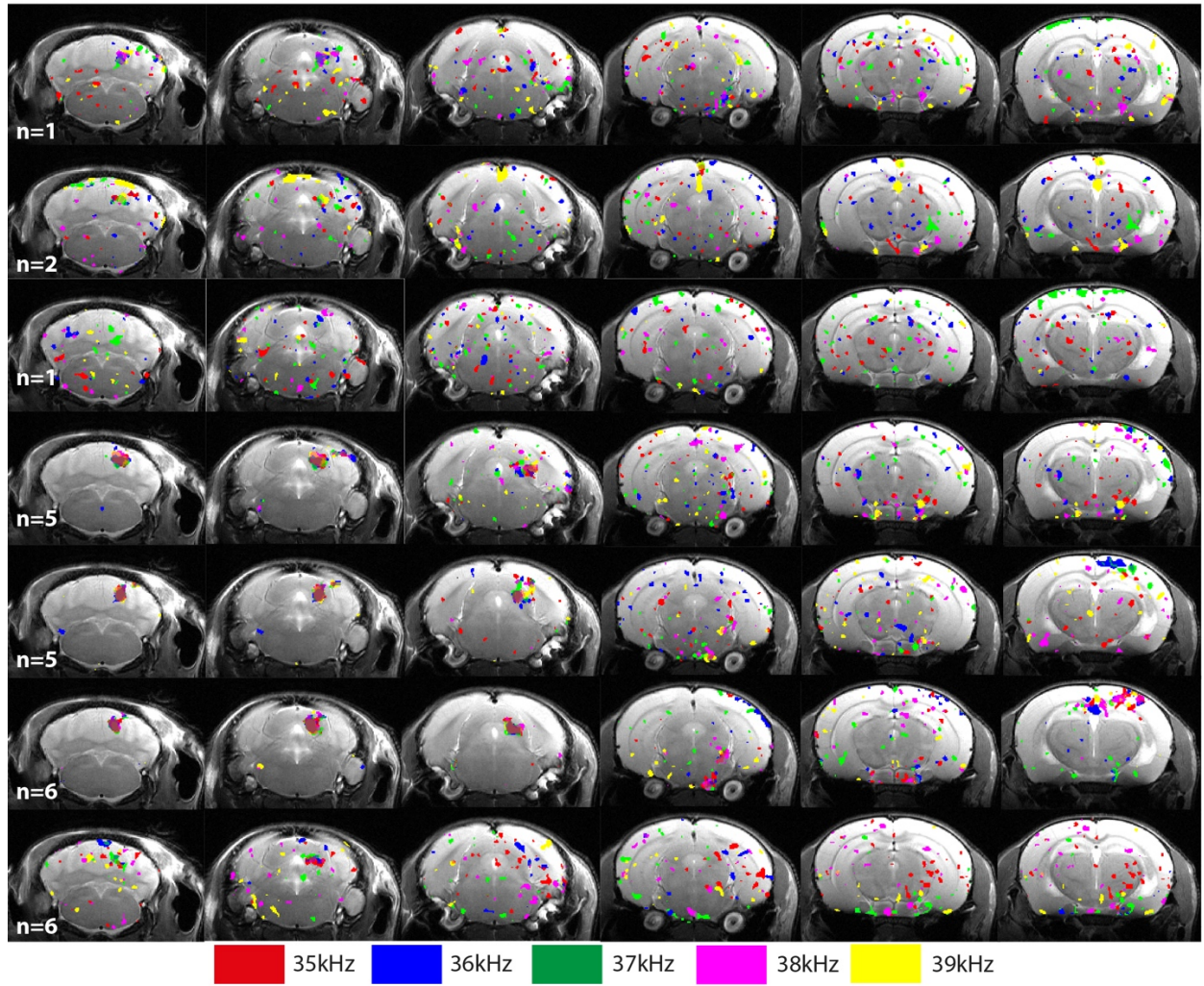

**Figure S6**. Individual 1kHz tonotopic maps per subject built from the top 3% positive t-values. Each subject shows a different pattern, only subjects with enough statistical power show IC and other auditory regions, and even in those cases all frequency responses are overlapped.

**Table 1.** Distribution of fMRI runs per animal, stimulation frequency and experiment. On the left, for experiment 1, each run consists of 5 stimulation blocks of the same frequency pure tone. On the right, for the additional experiment 2, each run consists of 5 blocks of different frequencies (either 35,36,37,38, or 39 kHz presented in a randomized order). The animal gender is indicated with "f" or "m".

| **Experiment 1** | **5 kHz** | **12 kHz** | **20 kHz** | **Experiment 2** | **35-39 kHz rnd** |
|---|---|---|---|---|---|
| Mouse 1 (m) | 1 | 1 | 1 | Mouse 6 (f) | 1 |
| Mouse 2 (m) | 1 | 1 | 1 | Mouse 7 (f) | 2 |
| Mouse 3 (m) | 2 | 2 | 2 | Mouse 8 (f) | 1 |
| Mouse 4 (m) | 1 | 1 | 1 | Mouse 9 (f) | 5 |
| Mouse 5 (m) | 1 | 1 | 1 | Mouse 10 (f) | 5 |
|  |  |  |  | Mouse 11 (f) | 6 |
|  |  |  |  | Mouse 12 (f) | 6 |

*1 run= 334 volumes, 5 stim. blocks of same frequency    *1 run= 334 volumes, 5 stim. blocks of different frequencies